\documentclass[twocolumn,showpacs,preprintnumbers,
amsmath,amssymb,aps,prd,nofootinbib,superscriptaddress,
eqsecnum]{revtex4}
\usepackage{graphicx}
\usepackage{dcolumn}
\makeatletter
\renewcommand{\p@subsection}{}
\makeatother

\newcommand{\Rmnum}[1]{\expandafter\@slowromancap\romannumeral #1@}
\usepackage[usenames]{color}

\newcommand{\be}{\begin{eqnarray}}
\newcommand{\ee}{\end{eqnarray}}

\def\lsim{\mathrel{\rlap{\lower3pt\hbox{\hskip1pt$\sim$}}
     \raise1pt\hbox{$<$}}} 
\def\gsim{\mathrel{\rlap{\lower3pt\hbox{\hskip1pt$\sim$}}
     \raise1pt\hbox{$>$}}} 

\def\la{\langle}
\def\ra{\rangle}

\begin{document}

\title{Half-Skyrmions and the Equation of State for Compact-Star Matter}

\author{Huan Dong}
\author{T.T.S. Kuo}
\affiliation{%
Department of Physics and Astronomy, Stony Brook University,
Stony Brook, NY 11794, USA}

\author{Hyun Kyu Lee}
\affiliation{%
Department of Physics, Hanyang University, Seoul 133-791, Korea
}

\author{R. Machleidt}
\affiliation{%
Department of Physics, University of Idaho, Moscow, ID 83843, USA
}

\author{Mannque Rho}

\affiliation{%
Institut de Physique Th\'eorique,
CEA Saclay, 91191 Gif-sur-Yvette c\'edex, France \& \\
Department of Physics, Hanyang University, Seoul 133-791, Korea
}

\date{\today}

\begin{abstract}
The half-skyrmions that appear in dense baryonic matter
when skyrmions are put on crystals modify drastically hadron
properties in dense medium and affect strongly the nuclear
tensor forces, thereby influencing the equation of state (EoS)
of dense nuclear and asymmetric nuclear matter. The matter comprised of
half skyrmions has vanishing quark condensate but non-vanishing
pion decay constant and could be interpreted as a hadronic dual of
strong-coupled quark matter. We infer from this observation {combined with certain predictions of hidden local symmetry in low-energy hadronic interactions }
a set of new scaling laws -- called ``new-BR" --  for  the parameters
in nuclear effective field theory controlled by
renormalization-group flow.
They are subjected to the EoS of symmetric and asymmetric nuclear
matter, and are then applied to  nuclear symmetry energies
 and  properties of compact stars.
 The changeover from the skyrmion matter to a half-skyrmion matter
that takes place after the cross-over density $n_{1/2}$
provides a simple and natural field theoretic explanation for the change of the EoS from soft to stiff at a density above that of nuclear matter required for compact stars as
massive as $\sim 2.4M_\odot$.  Cross-over density in the range
 $ 1.5n_0 \lsim n_{1/2} \lsim 2.0 n_0$ has
been employed, and the possible skyrmion half-skyrmion
coexistence {or cross-over} near $n_{1/2}$ is discussed.
 The novel structure of {the tensor forces and} the EoS obtained with the new-BR scaling is relevant for neutron-rich nuclei and compact star matter and
could be studied in RIB (rare isotope beam) machines.

\end{abstract}


\maketitle


\section{Introduction}
The topological soliton called skyrmion~\cite{skyrme} has turned
out to be exceedingly pervasive in a variety of space-time dimensions
ranging from 3 to 5 in many areas of physics~\cite{multifacet} and
has been beautifully observed in such systems as quantum Hall or
cold atoms and more recently in a monoatomic magnetic film
(see e.g., \cite{heinze}). In contrast, the situation with its role
in nuclear physics has been much less clear and with rather limited success.
In this note, we make an attempt to uncover the power, hitherto unexploited,
of skyrmion in strong interaction physics focusing on nuclear and dense
matter. In contrast to condensed matter, the effect of skyrmion structure
in strong interactions turns out to be indirect and hence much less
transparent. In this work we show with simple plausible arguments
that the skyrmion picture can indeed make a novel prediction
on the properties of compact stars that has not been made thus far
by other approaches.

The arguments made in formulating the theoretical framework
are neither rigorous nor completely unambiguous. { Although the crystal structure which is valid in the large $N_c$ limit could be applicable at very large density, it is not clear that it can be used in the density regime that we are concerned with, which will be a few times the normal nuclear matter density. What we will be exploiting is, however, the topological structure provided by the skyrmion configuration, which is insensitive to  spatial symmetry.}  In proceeding we will rely on what
Nature indicates at normal densities and then extrapolate to high
densities using a hidden local symmetry (HLS) structure with
well-defined degrees of freedom .

The starting point of our work is that when a large number of
skyrmions as baryons are put on an FCC (face-centered-cubic) crystal
to simulate dense
matter, the skyrmion matter undergoes a transition to a matter consisting
of half-skyrmions~\cite{goldhaber} in CC configuration at a density
that we shall denote as $n_{1/2}$. This density is difficult to pin
down precisely but it is more or less independent of the mass of the
dilaton scalar, the only low-energy degree of freedom that is not
well-known in free space. It has been estimated to lie typically at
between 1.3 and 2 times  the normal nuclear matter density $n_0$
~\cite{half}.

The half-skyrmion phase, made up of fractionized baryon numbers,
is characterized by the quark condensate $\la\bar{q}q\ra$ that vanishes
on the average in the unit cell with, however, chiral symmetry still
broken, so the pion is present. It likely has an inhomogeneous
spatial distribution of baryon density. There is no obvious order
parameter for the ``transition" although there can be higher-dimension
field operators representing an emergent symmetry that could be identified at quantum level. What can distinguish the two ``phases" are the different degrees of freedom
with different topological charges.

Among the predictions made so far with the half-skyrmion phase,
the most striking one  -- which is the main object of this article and has not been made by other approaches --  was that the presence of $n_{1/2}$ strongly modifies the tensor forces in nuclear interactions and in particular the symmetry energy at densities
$n> n_{1/2}$~\cite{LPR,LR12}.

In this note, we confront our predictions
with nature by translating the (semi-)classical results of \cite{LPR}
into the parameters of an effective Lagrangian having chiral symmetry
and conformal symmetry, and then do a quantum-EFT calculation for nuclear
matter and compact-star matter using a renormalization-group (RG) based formalism~\cite{kuo}.
For the range $n_{1/2}= (1.5-2.0) n_0$
considered in \cite{LPR}, we have applied our formalism to neutron star
calculations and a comparison of our results
 with   the recently discovered two-solar-mass neutron star
\cite{demorest2010} will be discussed. An interesting result of
the calculation is that the skyrmion-half-skyrmion crossover
makes the EoS stiffer at the crossover density $n_{1/2}$
 and beyond, thereby leading to more massive stars.

In a nutshell, our strategy is as follows. Up to the nuclear matter density $n_0$, our nuclear effective field theory (EFT) will be guided by symmetries of low-energy QCD (such as chiral symmetry, hidden local symmetry etc.) backed by nuclear phenomenology available up to density near $n_0$. There the effective Lagrangian will be endowed with parameters suitably scaling in the vacuum sliding with the density. We will assume that one can use the same EFT up to the density $n_{1/2}$ at which half-skyrmions appear  which we take to be above but not far above $n_0$. Above $n_{1/2}$ for which there are neither experimental data nor model-independent theoretical tools available, we will take the properties indicated by the skyrmion-half-skyrmion transition based on hidden local symmetric Lagrangian and certain predicted property of hidden local fields as the chiral critical point is approached. The effective Lagrangian so given is then translated into an effective nuclear field theory that is subject to many-body techniques that account for high order quantum effects.

\section{New-BR in the half-skyrmion phase}\label{newbr}
What plays a key role in our development is the nuclear symmetry
energy computed in \cite{LPR} in dense skyrmion matter. There it was
found that the symmetry energy $E_{sym}$ figuring in the energy
per particle of asymmetric nuclear matter at density $n$ in the form
\be
E(n,\alpha)=E(n,\alpha=0)+E_{sym}(n) \alpha^2
+{\cal O}(\alpha^4)\label{E}
\ee
where

$\alpha=(n_n-n_p)/(n_n+n_p)$ with $n_n(n_p)$ the number density of
neutrons (protons) is given by
\be
E_{sym}\approx\frac{1}{8\lambda_I}\label{sym}
\ee
where $\lambda_I$ is the isospin moment of inertia obtained
by rotational quantization of the multi-skyrmion system which is
given by an integral over the unit cell of a certain combination of
the skyrmion configuration. We should understand that this is a
quasi-classical potential energy contribution { {coming at ${\cal O}{(1/N_c)}$ in the large $N_c$ expansion}} and contains no kinetic
energy term. In what follows, we shall take into account quantum
corrections arising from nuclear correlations that are higher order in $1/N_c$. For the moment we focus on (\ref{sym}). A striking feature of (\ref{sym}) discovered
in \cite{LPR} is a cusp at $n_{1/2}$ of the symmetry energy which decreases from $n_0$ to $n_{1/2}$ and then increases for $n>n_{1/2}$.
 Now given the classical nature of (\ref{sym}) and the neglect of
the kinetic energy term, one cannot expect this feature to
appear unscathed in experiments. In order to confront nature,
one has to go beyond the classical approximation of the
skyrmion crystal. {How to systematically make quantum corrections within the skyrmion crystal approach is not yet known}. What we shall instead do is to ``translate" the classical result of \cite{LPR} into the framework of an effective field theory
treated at mean-field of a HLS Lagrangian~\cite{HY:PR}
that contains all relevant degrees of freedom at the energy
scale involved, i.e., baryons, pions and vector mesons. {There is of course a certain arbitrariness in doing this but we shall rely on what Nature indicates.}
In addition, a dilaton scalar denoted $\chi$ is introduced to account for the
spontaneously broken conformal symmetry as precisely
defined in \cite{LRdilaton}. The work in \cite{LPR} uses the nonlinear
sigma model, involving only pions and baryons (emerging as skyrmions).
However the nonlinear sigma model Lagrangian is gauge-equivalent to
the HLS Lagrangian, hence can capture the physics
of vector mesons as was proposed in \cite{BR91}. Our strategy
which is consistent with the spirit of the renormalization group is then
 to  do effective field theory calculation with
this Lagrangian, with the parameters of which ``running" with the
{\it intrinsic} medium dependence as formulated in \cite{BR91}.
We will refer to this medium dependence as
``BR scaling."

We will now describe how the cusp structure in (\ref{sym})
can be reproduced by an effective Lagrangian in mean field.

{Up to the density $n_{1/2}$, our effective Lagrangian will
carry the parameters scaling as introduced in \cite{BR91}.
Let us call it ``old-BR." They are of the form
\be
m_V^*/m_V \approx m_N^*/m_N\approx
f_\pi^*/f_\pi\equiv \Phi_{I}\label{oldBR}
\ee
and
\be
g_V^*/g_V&\approx& 1,\label{oldBR1}
 \ee
where the asterisk represents density dependence, $f_\pi$ is the
pion decay constant, the subscripts $N$ and $V$ stand, respectively,
for the nucleon and the vector mesons $V=\rho, \omega$\footnote{Whenever necessary, as will be the case for $n>n_{1/2}$, we will specify whether it is $\rho$ or $\omega$} and $g_V$ is the hidden gauge coupling constant $g$ standing for both $V=\rho, \omega$~\cite{BR:DD}.
It has been assumed~\cite{BR91,BR:DD} that the flavor $U(2)$ symmetry applies to $(\rho, \omega)$ in baryonic matter up to the normal nuclear matter density $n_0$ and will be assumed in what follows, up to $n_{1/2}$ as it does in matter-free space.
However at
$n\gsim n_{1/2}$, the fractionization of the skyrmions produces
a change in the intrinsic scaling as~\cite{LPR}
\be
m_\rho^*/m_\rho \equiv \Phi_{II}^\rho,\ \
m_N^*/m_N \equiv \Phi_{II}^N =y(n)\label{newBR}
\ee
where $y(n)$ is an order 1 constant that is more or
less density-independent as explained below. The scaling
$\Phi_{II}^{\rho}$ is unknown except
(perhaps) very near chiral transition.
It needs not scale in the same way as  $\Phi_I$ does as explained below.
Very near chiral transition at $n=n_c$, however, the HLS
theory has, whether viable or not, a definite
prediction thanks to the ``vector manifestation fixed point (VM)"
at which the matching of both the vector and axial-vector
correlators gives~\cite{HY:PR}
\be
 m_\rho^*/m_\rho\approx g_\rho^*/g_\rho\rightarrow \la\bar{q}q\ra^*/\la\bar{q}q\ra,\ \ n\rightarrow n_c
\label{gscaling}
\ee
where $q$ stands for chiral quark field. Unless we assume that $U(2)$ symmetry holds in medium -- that we will not as explained below, the hidden local symmetry argument does not give any prediction as to how the $\omega$ mass and the $\omega$-NN coupling behave in medium for $n>n_{1/2}$. For simplicity, we will simply take
\be
 m_\omega^*/m_\omega\approx m_\rho^*/m_\rho.
\label{omegascaling}
\ee
As for the $\omega$-NN coupling $\equiv g_\omega$, it is very much subtler and we will specify it later. We will call (\ref{newBR})-(\ref{omegascaling}) ``new-BR." The difference from the old-BR is lodged in the density regime $n\geq n_{1/2}$.}

There are two points to note here:
\begin{itemize}
\item
One is that at $n_{1/2}$ the scaling parameter changes from the pion decay constant scaling as $(f_\pi^*/f_\pi)^2 \sim {\la\bar{q}q\ra^*/\la\bar{q}q\ra}$ to the hidden local symmetry coupling constant $g_\rho$ scaling linearly as in (\ref{gscaling}). This changeover was already observed in \cite{BR:DD} from phenomenology. Since $g_\rho$ is directly connected, via renormalization group flow, to the quark condensate which is the bona-fide order parameter of chiral symmetry in the chiral limit, near the VM fixed point, it is the vector meson mass $\propto g_\rho$ that carries information on chiral symmetry, not the pion decay constant. This changeover of the scaling from $f_\pi^*$ to $g_\rho^*$ that accounts for that $\Phi_I$ and $\Phi_{II}$ need not have the same behavior in density is reflected in the half-skyrmion phase in that the pion decay constant drops only slowly in contrast to the quark condensate which drops to zero at $n_{1/2}$. It is important to note that the hidden gauge coupling $g_\rho$ scales in the same way as the $\rho$-meson mass does at high density whereas at low density up to $n_{1/2}$, the gauge coupling stays  unscaling. This difference will turn out to have a drastic effect on the $\rho$ tensor force for $n>n_{1/2}$
\item
The second point, also connected to the slowly dropping pion decay constant, is that the nucleon mass scales little beyond $n_{1/2}$, remaining non-zero at the chiral transition. This resembles -- and we believe is related to -- the nucleon mass in the parity-doublet nucleon model where there is a rather large chirally invariant mass $m_0$ that remains at the transition~\cite{detar}. In our application to be given below, we will consider $m_0\sim (0.7-0.8)m_N$.
\end{itemize}

That the new-BR affects the nuclear tensor forces across the density $n_{1/2}$ was explained in \cite{LPR,LR12}. So we will skip the details and briefly summarize only the main features that we will need below.

If one takes the nucleon to be heavy while other hadrons, i.e., mesons, are light, then one can take the nonrelativistic approximation for the nucleons and write the effective tensor forces in medium in the usual form with the parameters of the Lagrangian carrying the intrinsic density dependence \`a la BR scaling. The two tensor forces contributed by the pion exchange and the $\rho$ exchange are given in the standard form with the masses and coupling constants replaced by the starred quantities. Dividing by the vacuum quantity $(C_M)^2=(\frac{f_{MN}}{4\pi})^2$ and writing $x_M^*=m_M^* r$,  we have the in-medium $\pi$ and $\rho$ tensor forces in the form
 \begin{eqnarray}
 V_M^T(r)/(C_M)^2 = S_M \tau_1 \cdot \tau_2 S_{12} (R_M^*)^2 m_M^\star Y(x_M^*)
\label{tenforce}
\end{eqnarray}
with
\be
Y(x_M^*)=\left[ \frac{1}{(x_M^*)^3} + \frac{1}{(x_M^*)^2}
+ \frac{1}{3 x_M^*} \right] e^{-x_M^*}
\ee
where $M=\pi, \rho$, $S_{\rho(\pi)}=+1(-1)$
and $R_M^*=f_{MN}^*/f_{MN}$. The crucial (very well-known) feature to note is that the two forces come with an opposite sign.

As argued in \cite{LPR}, the pion tensor force can be taken unscaling in all relevant density range. In fact, one can verify explicitly that using suitably scaling parameters for all parameters that enter in the pion tensor force, such as $f_\pi$ etc., gives results that are close to those obtained by taking {\em all} the parameters unscaled~\cite{LR12}. Thus $R_\pi^*\approx 1$ and $m_\pi^*\approx m_\pi$ in Eq.~(\ref{tenforce}).
As for the $\rho$ tensor, what remains to be determined is the scaling of $R_\rho^*$. It follows straightforwardly from (\ref{oldBR}) and (\ref{newBR}) that
\be
R_\rho^*&=& \frac {g_V^* m_V^* m_N}{g_V m_V m_N^*}\nonumber\\
&\approx& \frac{g_V^*}{g_V}\approx 1\ \ {\rm for}\ \ \ 0\lsim n \lsim n_{1/2}\label{R1}\,
\ee
{ and
\be
R_\rho^*&=& \frac {g_\rho^* m_\rho^* m_N}{g_\rho m_\rho m_N^*} \approx \frac{g_\rho^*}{g_\rho}\Phi_{II}^\rho/y(n)\nonumber\\
& \approx&(\Phi_{II}^\rho)^2/y(n)\ \ {\rm for} \ \  n_{1/2}< n \lsim n_c\label{R2}
\ee
where $n_c$ is the putative chiral transition density. Since $g_\rho^*/g_\rho\sim \Phi_{II}^\rho$ in Region II, we have (\ref{R2}) with $R_\rho^*$ scaling as $\sim  (\Phi_{II}^\rho)^2$ for $n\gsim n_{1/2}$} and this makes a big change in the behavior of the net tensor force. Up to density $n\approx n_0$, the scaling $\Phi_I$ makes the total tensor strength weakened at increasing density because the increased $\rho$ tensor eats into the pion tensor~\cite{BR:tensor,LR12}. Recently this mechanism has been shown to explain the long standing problem of the carbon 14 dating~\cite{c14} which in turn determines how $\Phi_I$ scales up to $n_0$.\footnote{It has been suggested that this could also be explained by certain short-range three-body forces~\cite{holt}. This of course does not mean that three-body forces are an {\it alternative} to the scaling mechanism. More on this point in Discussions section.} If the scaling $\Phi_I$ continued beyond $n_0$, then it would make the net tensor attraction vanish at $n\sim 2n_0$~\cite{LR12}. Now with the new scaling, this behavior no longer holds. The simple prediction is that the net tensor-force strength will cease to drop at $n_{1/2}$. Just to have an idea of what this does, take $\Phi_{II}\approx \Phi_I$. A simple estimation shows that when density reaches $n\sim (2-3)n_0$, the $\rho$ tensor becomes totally negligible. What remains is only the pion tensor. When this happens, $\pi^0$'s could condense in a crystalline form as suggested in \cite{LPR,LR12}.

We now argue that an effective field theory at mean field with the tensor force that follows from the new-BR can reproduce the cusp structure in the symmetry energy (\ref{sym}) seen in the skyrmion-crystal calculation. This can be seen from the fact that the symmetry energy is dominated by the tensor force~\cite{brown-machleidt,tensordominance}. A simple formula that captures the essential physics of the tensor forces is that of Brown and Machleidt~\cite{brown-machleidt} that we rewrite including the new-BR,
\be
E_{sym}\approx \frac{C}{\bar{E}}\la V_T^2\ra\label{E-tensor}
\ee
where $C$ is a known constant, $\bar{E}$ is the average energy appropriate for the tensor force, $\approx 200$ MeV, $V_T$ is the radial part of the tensor force that includes the effect of the new-BR. There is again the kinetic term which we will ignore as before. In the form of (\ref{E-tensor}), the cusp structure then follows immediately from the discussions given above for the behavior of the tensor forces across $n_{1/2}$, i.e., the decrease of the net tensor force strength from $n_0$ to $n_{1/2}$ and its increase after $n_{1/2}$ with the pion tensor taking over the strength. We take this as a support for the skyrmion crystal - mean-field EFT transcription.

We will see below how nuclear correlations that go beyond the mean-field approximation modify this cusp structure. It actually smoothes it without completely eliminating it. The effective field theory anchored on the Lagrangian endowed with the new-BR is applied first to nuclear matter and then to compact-star matter addressing the issue of the maximum neutron-star mass vs. radius.
\section{Nuclear Equation of State}\label{eos}
So far we have been discussing qualitative features impacted by the new-BR. We now confront quantitatively the scaling relations (\ref{oldBR}),  (\ref{newBR}), (\ref{R1}) and (\ref{R2}) with the properties of symmetric as well as asymmetric nuclear matter. We  incorporate the new-BR in the nuclear effective field theory in which the RG-implemented $V_{low-k}$ plays a key role. More specifically, we apply the new-BR scalings discussed in Section \ref{newbr} to nuclear matter, both symmetric and asymmetric,  and to the nuclear symmetry energy. The special features in the new-BR are
the scaling of the nucleon mass $y(n)$ and that of the vector coupling $g$. As we shall elaborate in the discussion section, these features are thought to be closely connected to how (most of) the nucleon mass is generated in the strong interactions.

Before continuing, let us concisely recapitulate how the new-BR
enters into an RG-implemented EFT. As argued in \cite{BR:DD}, it
involves two decimations in the RG sense. Starting with an effective
chiral Lagrangian, one first decimates in matter-free space from
the chiral scale $\Lambda_\chi\sim 4\pi f_\pi\sim 1$ GeV down to
the first decimation scale  $\Lambda\sim 3$ fm$^{-1}$. What results
is the $V_{low-k}$ that is used in our calculation. Then doing
many-body calculations for nuclear systems with the parameters of
$V_{low-k}$ running \`a la new-BR amounts to doing the second
decimation. In fact this second decimation is equivalent to doing
 a Landau Fermi-liquid theory calculation as formulated
in ~\cite{BR:DD}. In doing this, we are ignoring 3-body
and higher-body forces. One should however recognize that part of
many-body force effects are embedded in the new-BR. One can think
of this as a sort of duality between the two as will be elaborated  later.

To suitably take into account the features mentioned above into a high-order effective field theory calculation, we shall carry out our calculations using the realistic
BonnS potential \cite{bonns}; this potential  is an extension of the
one-boson-exchange BonnA potential \cite{mach89} with the provision that
the nucleon and meson mass as well as the vector coupling be scaled \`a la new-BR.
As discussed in Section \ref{newbr}, we employ the following two-region
scalings characterized by the transition densities $n_{1/2}$ and $n_c$
(respectively for the skyrmion-half-skyrmion and chiral transitions).
For density  $0<n<n_{1/2}$ (Region-I), we use\footnote{We must stress that except for low density $\lsim n_0$ (and possibly high density near
the chiral transition point in the chiral limit as predicted in HLS),
the precise form of the scaling is not known, so what we take
should be understood as more
of a convenient parametrization guided, whenever feasible, by
phenomenology. Furthermore there is nothing that suggests that the
scaling should be identical for all mesons. }
\begin{equation}
\frac{m_M^*}{m_M}=\frac{m_N^*}{m_N}=\Phi_I(n);
\Phi_I(n)=\frac{1}{1+c_I\frac{n}{n_0}}
\end{equation}
and {
\be
R_\rho^*=1.
\ee }
In the above $M=(V,S)$ and $N$ stand respectively for meson (both vector and scalar) and nucleon.
For density  $n_{1/2}<n<n_c$ (Region-II), we use
\begin{equation}
\frac{m_M^*}{m_M}=\Phi_{II}^M(n);
\Phi_{II}^M(n)=\frac{1}{1+c_{II}\frac{n}{n_0}}
\end{equation}
for mesons and
\begin{equation}
\frac{m_N^*}{m_N}=\Phi_{II}^N(n)=y(n)
\end{equation}
for nucleons. {We use the $R^*$ scaling in II as
\be
R_\rho^*&=& \frac{g_\rho^*}{g_\rho}\Phi_{II}^M (n)/y(n)= (\Phi_{II}^M)^2/y(n).
\label{correctR}
\ee}
The above scaling functions $\Phi_I$ and $\Phi_{II}$ are
in general not continuous at the boundary density $n_{1/2}$.
This discontinuity may be a mere artifact of the simplification
we are adopting. In the present
work, as to be discussed later, we shall choose the parameters
contained in them so that these two functions are nearly
continuous (to avoid drastic discontinuity)
 at $n_{1/2}$. In addition, we shall employ
 two Fermi-Dirac
functions to smoothly join the scaling functions $\Phi _I$ and $\Phi _{II}$
so that the resulting scaling function $\Phi$ is ensured to be
 continuous  at the boundary. A similar procedure will also be employed
for the $R^*$ scalings in the two regions. To illustrate, the smoothed
scaling function $\Phi$ is constructed as
\begin{equation}
 \Phi=F_{<}(n_{1/2})\Phi_I +F_{>}(n_{1/2})\Phi_{II},
\end{equation}
with
\begin{eqnarray}
   F_{<}(n_{1/2})&=&[1+e^{(n-n_{1/2})/\delta}]^{-1}, \nonumber \\
   F_{>}(n_{1/2})&=&[1+e^{(n_{1/2}-n)/\delta}]^{-1}
\end{eqnarray}
where $\delta$ is a smoothness parameter.
In the present work, we shall use $\delta/n_0 \simeq$ 0.05-0.10.
It turns out that within this range our results are satisfactorily stable
with respect to $\delta$.

\vskip 0.3cm
We have adopted the following procedure for choosing
the  parameters of the above scaling functions. First  we require
the parameters in Region I so that they
 satisfactorily reproduce the empirical nuclear matter saturation
properties (saturation density  $n_0\simeq 0.16$ fm$^{-3}$
and average energy per nucleon  $E_0/A\simeq -16$ MeV at saturation).
The choice for the parameters in Region II will be addressed later.
We shall calculate $n_0$ and $E_0/A$ using a low-momentum
ring-diagram approach
\cite{siu09,dong09,siu08,dong10,kuo}, where
the $pphh$ ring diagrams are summed to all orders
within a  model space of decimation scale $\Lambda$.
Few low-order (1st-, fourth- and eighth-order) such diagrams
are displayed in Fig.~1. Note that each vertex of the diagrams
is a low-momentum interaction $V_{low-k}$ which
 is obtained from a realistic NN potential $V_{NN}$ using
 a renormalization group approach where the momentum components beyond
a decimation scale $\Lambda$ are integrated out
\cite{bogner01,bogner02,bogner03,bogner03b}.

\begin{figure}[here]
\scalebox{0.4}{\includegraphics{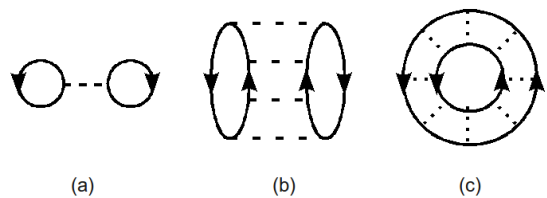}}
\caption{Diagrams included in the all-order \emph{pphh} ring-diagram
summation for the ground state energy  of nuclear matter.
Each dashed line represents a $V_{low-k}$ vertex.}\label{figure1}
\end{figure}

More precisely $V_{low-k}$
is given by the following $T$-matrix equivalence equations:
\begin{multline}
T(k',k,k^2)=V_{\rm NN}(k',k) \\
+\frac{2}{\pi}\mathcal{P}\int_0^\infty
\frac{V_{\rm NN}(k',q)T(q,k,k^2)}{k^2-q^2}q^2dq,
\end{multline}
\begin{multline}
T_{\rm low-k}(k',k,k^2)=V_{\rm low-k}(k',k)\\
+\frac{2}{\pi}\mathcal{P}\int_0^\Lambda
\frac{V_{\rm low-k}(k',q)T_{\rm low-k}(q,k,k^2)}{k^2-q^2}q^2dq,
\end{multline}
\begin{equation}
 T(k',k,k^2)=T_{low-k}(k',k,k^2); (k',k) \leq \Lambda.
\end{equation}
In the present work the above $V_{NN}$ is chosen to be the
 realistic BonnS \cite{bonns} NN interaction. (The new-BR
 scalings we have established above enter into the  meson parameters as well as the nucleon mass of this potential with the varying density.)
 $\mathcal{P}$ denotes principal-value integration and the intermediate state momentum \emph{q} is integrated
from 0 to $\infty$ for  the whole-space $T$ and from 0 to $\Lambda$
for $T_{\rm low-k}$.
 Because we shall calculate the nuclear symmetry energy
$E_{sym}(n)$ up to $n\sim5n_0$, we shall use
$\Lambda=3$ fm$^{-1}$ \cite{kuo}.
 The above $V_{low-k}$ preserves the low-energy
phase shifts { in the vacuum} (up to energy $\Lambda^2$) and the deuteron binding energy of $V_{NN}$. (For example, the deuteron binding energy given by
$V_{low-k}$ of $\Lambda=$ 2.0 and 3.0 fm$^{-1}$ are both -2.226 MeV.)
Since $V_{low-k}$ is obtained by integrating out
the high-momentum
components of $V_{NN}$, it is a smooth `tamed' potential which is suitable
for being used directly in many-body calculations.
The familiar HF approximation for nuclear matter corresponds
to the inclusion of only the first-order diagram (a) of the figure.
In contrast, the $pphh$ ring diagrams such as those shown in Fig.~\ref{figure1} are included to all orders in our nuclear matter calculations.

With such ring diagrams summed to all orders \cite{siu09,dong09},
 the ground-state energy of asymmetric nuclear matter is expressed as
$E(n,\alpha)=E^{free}(n,\alpha)+ \Delta E(n,\alpha)$
where $E^{free}$ denotes the energy for the non-interacting system
and $\Delta E$, the energy shift due to the NN interaction,
is given by the all-order sum of the \emph{pphh} ring diagrams as illustrated
in Fig.~\ref{figure1}. We include in general three types of ring diagrams, the
proton-proton, neutron-neutron and proton-neutron ones.
 The proton and neutron
Fermi momenta are, respectively,
$k_{Fp}=(3\pi^2n_p)^{1/3}$
and $k_{Fn}=(3\pi^2n_n)^{1/3}$, where $n_p$ and $n_n$ denote respectively
the proton- and neutron-density. The asymmetric parameter is
$\alpha \equiv (n_n-n_p)/(n_n+n_p)$.
With such ring diagrams summed to all orders,  we have
\begin{multline}\label{eng}
\Delta E(n,\alpha)=\int_0^1 d\lambda
\sum_m \sum_{ijkl<\Lambda}Y_m(ij,\lambda) \\ \times Y_m^*(kl,\lambda) \langle
ij|V_{\rm low-k}|kl \rangle,
\end{multline}
where the transition amplitudes $Y$ are obtaind from a $pphh$ RPA equation
\cite{siu09,dong09}.
Note that $\lambda$ is a strength parameter,
integrated from 0 to 1. The above ring-diagram method reduces to the
usual HF method if only the first-order ring diagram
is included. In this case, the above energy shift becomes
$\Delta E(n,\alpha)_{HF}=\frac{1}{2}
\sum n_i n_j\langle ij|V_{\rm low-k}|ij \rangle$ where
$n_k$=(1,0) if $k(\leq,>)k_{Fp}$ for  proton
and $n_k$=(1,0) if $k(\leq,>)k_{Fn}$ for  neutron.

 The above $V_{low-k}$ ring-diagram framework has been applied to
symmetric and asymmetric nuclear matter \cite{siu09,dong09}
 and to the nuclear symmetry energy
\cite{kuo}.  This framework has also been tested by  applying it  to dilute cold neutron matter in the  limit that the $^1S_0$ scattering
length of the underlying interaction approaches infinity
\cite{siu08,dong10}. This limit -- which is a conformal fixed point --  is usually referred to as the unitary limit, and the corresponding potentials the unitarity potentials.
 For many-body systems at this limit, the ratio
$\xi \equiv E_0/E_0^{free}$ is
expected to be a universal constant of value $\sim 0.44$. ($E_0$ and $
E_0^{free}$
are, respectively, the interacting and non-interacting ground-state
energies of the many-body system.)  The above ring-diagram method has been
used to calculate
neutron matter using several very different unitarity potentials
(a unitarity CDBonn potential  obtained by tuning
its meson parameters,
 and several square-well unitarity potentials) \cite{siu08,dong10}.
The $\xi$ ratios
given by our calculations for all these different unitarity potentials are
all close to 0.44, in good agreement with the Quantum-Monte-Carlo results
(see \cite{dong10} and references quoted therein).
In fact our ring-diagram results for $\xi$ are significantly better than
those given by HF and BHF (Brueckner HF) \cite{siu08,dong10}. It is
 desirable that the above unitary calculations have provided
satisfactory results,  supporting the reliability of
 our $V_{low-k}$ ring-diagram framework for calculating the nuclear matter EoSs.
\section{Results}
We recall that the new-BR has an assumption which is not an
immediate consequence of chiral symmetry. Specifically the premise
of the vector manifestation associated with hidden local symmetry
states that as one approaches the VM fixed point,
$m_\rho^*/m_\rho\rightarrow g_\rho^*/g_\rho\rightarrow 0$, which is not dictated
by chiral symmetry alone. Here we will take the point of view that
the vector manifestation property is operative after the
half-skyrmion onset density $n_{1/2}$.
\begin{figure}[hbt]
\scalebox{0.34}{\includegraphics[angle=-90]{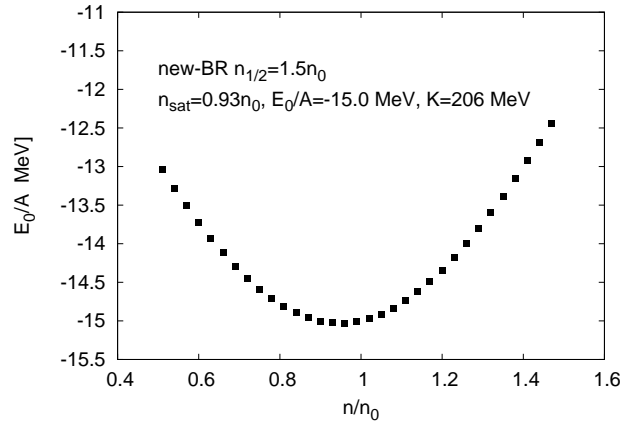}}
\caption{ New-BR EoS for symmetric nuclear matter
calculated with $n_{1/2}=1.5n_0$.
See text for more explanations.}\label{figure2}
\end{figure}

 Let us first consider the EoS of symmetric nuclear matter in the low
density region ($n \lsim n_{1/2}$), the main purpose here being the choice of the
$c_I$ parameters so that the empirical saturation properties of
symmetric nuclear matter are satisfactorily reproduced.
In Fig.~\ref{figure2}, we present  our results for
symmetric nuclear matter calculated with parameters $c_I$=0.130
for nucleon and $\rho$-meson, =0.121 for $\sigma$-meson and =0.139
for $\omega$-meson.\footnote{Here we are doing some
fine-tuning for a better fit but the small differences in $c_I$'s are of course of no significant meaning.}
The EoS of this figure gives
 ground-state energy per nucleon   $E_0/A$ =-15 MeV,
 saturation density
$n_{sat}=0.93 n_0$ and compression modulus $K$= 208 MeV, all in
satisfactory agreement with the empirical values.
  (Here and in Fig.2 $E_0/A$ is the same as $(E(n, \alpha=0)-m_N)$
of Eq.(2.1).)
 The above calculation has employed $n_{1/2}=1.5n_0$. As to be
presented later, we have also carried out calculations with $n_{1/2}=2n_0$
and the saturation properties given by them are nearly
the same as the $n_{1/2}=1.5n_0$ case.  Recall that a decimation scale
of $\Lambda=3$~fm$^{-1}$ has been employed in the
above calculation, and it will be used in what follows.

As discussed in \cite{siu09,dong09}, the use of realistic $V_{NN}$
with the old-BR~\cite{BR91,hatsuda,rapp} leads to satisfactory
nuclear matter saturation properties. As is seem in  Fig. 2,
the new-BR does also lead to satisfactory
nuclear matter saturation properties,  even though
these two scalings are different for $n\gsim n_{1/2}$.
The main differences between them
are in the scaling of the nucleon mass and the
HLS coupling $g$.  While the nucleon mass does scale in Region-I
with the change in $\la\bar{q}q\ra^*$, its scaling more or less
stops at $y(n_{1/2})$ for $n>n_{1/2}$ and is assumed to change
drastically only at $n_c$. The gauge coupling $g$ on the other
hand remains unchanged up to $n_{1/2}$ and drops roughly proportional
to $\la\bar{q}q\ra^*$ afterwards as suggested in \cite{BR:DD}.

  Before proceeding to the EoS for $n\gsim n_{1/2}$, let us first
discuss the scaling parameters we have employed.
 The scaling functions we have used in Region I ((3.1-3.2))
are similar to those employed in the
 old-BR \cite{BR91,hatsuda} and Ericson (ER) \cite{ericson} scalings.
The ER scaling is based on the quark condensate relation \cite{ericson}
\begin{equation}
\frac{\langle\bar{q}q\rangle^*} {\langle\bar{q}q\rangle}
=\frac{1}{1+\frac{n \Sigma_{\pi N}}{f_\pi^2m_\pi^2}},
\end{equation}
where $\Sigma _{\pi N}$=45$\pm7$ MeV \cite{gasser}.
Then the ER scaling for hadrons in medium reads
\begin{equation}
\frac{m^*}{m}=\left(\frac{1}{1+D\frac{n}{n_0}}\right)^{1/3}
\end{equation}
with $D=\frac{n_0 \Sigma_{\pi N}}{f_\pi^2m_\pi^2}$.
Using the empirical values for
($\Sigma _{\pi N},~n_0,~f_{\pi},~m_{\pi}$), we have
\emph{D}=  0.35$\pm$0.06.
Note that for the low-density region this relation agrees well with the
 parametrization for the old-BR~\cite{BR91,hatsuda}
\begin{equation}
     \frac{m^*}{m}=1-C\frac{n}{n_0}
\end{equation}
where \emph{C} is a constant of value $\sim 0.15$.

It may be noted that our new-BR in Region I ((3.1-3.2)) is
consistent
with the above Ericson
 scaling in the $n<n_{1/2}$ region if the $c_I$ scaling parameters
are chosen to have values
near $D/3 \simeq 0.12 \pm 0.02$.
It is encouraging that the $c_I$ parameters
we have employed so as to give satisfactory nuclear matter saturation
properties (Fig.~\ref{figure2})
are indeed quite close to the value of $D/3$ given by QCD theories.

\begin{figure}[here]
\scalebox{0.34}{\includegraphics[angle=-90]{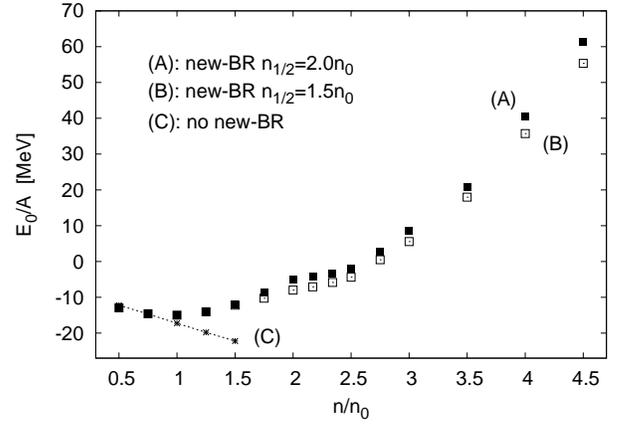}}
\caption{Comparison of the EoS for symmetric nuclear matter calculated with
$n_{1/2}=2.0$  (solid square) and $1.5n_0$ (open square).
 See text for more explanations.}\label{figure3}
\end{figure}

  We now consider the EoS for $n>n_{1/2}$. In Fig. 3 we present results
for two choices for the half-skyrmion onset densities, namely
$n_{1/2}$= 2.0 and 1.5$n_0$.
In addition we also present the EoS (labelled (C) in the figure)
 obtained with the unscaled BonnS \cite{bonns} potential.
As seen this EoS does not have  satisfactory nuclear matter saturation
properties; it would give saturation density much higher
than the empirical value of $\sim 0.16 fm^{-3}$ as well as
a saturation energy much lower than the empirical value
of $\sim -16$ MeV. In contrast, the new-BR EoSs (A) and (B),
respectively for
$n_{1/2}$= 2.0 and 1.5$n_0$,
both have satisfactory  saturation properties.
In calculating (A) and (B), we have used the same  $c_I$ parameters as
listed earlier. Thus  (A) and (B) are equivalent for $n<1.5n_0$,
 both having the same saturation properties ($E_0/A$=-15 MeV,
$n_{sat}=0.93 n_0$ and  $K$= 208 MeV).

Turning to the EoS in Region II ($n>n_{1/2}$), we note from
(\ref{R1}) and (\ref{R2}) (or (3.1-3.5)) that the scalings for $R_\rho^*$  controlling
the $\rho$ tensor force in Region II are significantly different
from those in Region I:
Other components of vector-meson-exchange nuclear forces are governed,
apart from the mass scaling,  by the scaling of the
hidden gauge coupling constant{
\be
g_\rho^*/g_\rho \approx g_\omega^*/g_\omega &\approx&1,~~n<n_{1/2};  \\
 g_\rho^*/g_\rho&\approx & [\Phi_{II}^M],~~ n>n_{1/2}.\label{g}
\ee }
Were the flavor $U(2)$ symmetry operative in Regin II,
the scaling (\ref{g}) would hold for both $\rho$ and $\omega$.
It turns out, however, that if the $\omega$-nucleon coupling
dropped in the same way as the $\rho$-nucleon coupling,
 nuclear systems would collapse in that region. We have found
that the repulsion provided by the $\omega$-exchange potential
is sensitively dependent on
 the $\omega$-nucleon coupling constant, and  a moderate dropping of this constant can drastically suppress the repulsion, making the system unstable at high densities {unless the nucleon mass dropped appreciably, which we do not consider realistic.} This signals that the coupling constant $g^*$ must be asymmetric in high density or higher members of the $\omega$ mesons in the infinite tower in holographic
QCD models that arise in string theory~\cite{SS} could intervene
in providing the necessary repulsion. In our calculation, we will
take $g_\omega^*/g_\omega\approx 1$  for  both Regions I and II.

Another difference is that the scaling of the nucleon
mass ($m_N^*/m_N$) is density dependent in Region I as is seen
in experiments while it is equal to a constant or slowly varying
in Region II. These differences can make the EoSs in these two
regions significantly discontinuous at $n=n_{1/2}$. As mentioned above,
this discontinuity could be an artifact of our schematic
treatment of the skyrmion-half-skyrmion transition. We have found that
this discontinuity can be made small  by suitably choosing the
scaling parameters in Region II. We have done so,
and for the EoSs (A) and (B) presented
in Fig. 3 we have used  $c_{II}=c_I$ for both (A) and (B),
 with y(n)=0.77 and $n_{1/2}=2.0n_0$ for (A),
and y(n)=0.78 and $n_{1/2}=1.5n_0$ for (B). The use
of the above $y(n)$ values is to have the $n<n_{1/2}$  energy curves
join smoothly with, respectively, their $n>n_{1/2}$ counter parts
at  $n_{1/2}$. It may be noted that  both $y(n)$ values are close to 0.80.
These parameter choices will be referred to respectively as A-parameters
and B-parameters. They will be used and tested in other calculations
such as nuclear symmetry energies and neutron stars later on.
{That the behaviors of $m_N^*$ and $g_\omega^*$ may be strongly
correlated in Region II will be discussed in the discussion section.
In view of the almost complete absence of model-independent
theoretical tools for these quantities in Region II, our strategy
will then be that the available heavy-ion experiments that probe
densities up to $\sim 4.5n_0$ give constraints on those parameters.
Calculating those parameters from the given theoretical framework remains to be done.}

\begin{figure}[here]
\scalebox{0.34}{\includegraphics[angle=-90]{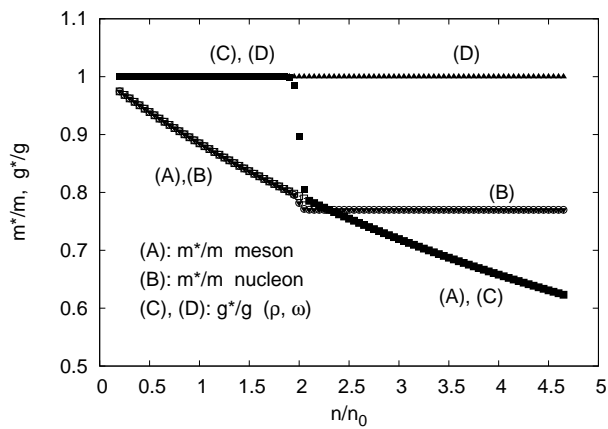}}
\caption{Plot of new-BR scalings for Regions I and II.
 See text for more explanations. }\label{figure4}
\end{figure}

It may be useful now to have a summary of the new-BR scalings employed in our present calculations. For this purpose, we present a plot of our $m^*/m$ and $g^*/g$ scalings, for the case of  $n_{1/2}=2n_0$, in Fig.~\ref{figure4}. As shown by line (A) there , the scalings for the $\rho$, $\omega$ and $\sigma$ masses are the same in both regions, recalling that for them we have $c_I=c_{II}$ and they are all close
to 0.13. (This value is used for plotting them in  the figure.)
As shown by line (B), the scaling of the nucleon mass  in Region I
is the same as the above mesons, but in region II it is equal to a constant $y(n)$.
($y(n)=0.77$ is used in the figure.) From (C) and (D), we see that
the scaling $g^*/g$  for $\rho$  is equal to one (i.e. unscaled)
in  region I and equal to the above meson scaling in Region II,
while the scaling  for $\omega$ is equal to one  in both regions.

As stated above, the above summary represents scalings of the intrinsic parameters of the underlying Lagrangian with appropriate symmetries (here, hidden local symmetry that captures the physics of vector mesons) with which our nuclear EFT is constructed.  In physical quantities,  the sharpness in changeover would be smoothed by many-body correlations as we find in the results.

 It is of interest that the EoSs (A) and (B) of Fig. 3 both exhibit  a narrow
 `plateau-like'  segment near $n\sim 2.2n_0$.
The occurrence of this  plateau-like  structure may indicate a
skyrmion-half-skyrmion transformation which is of interest for
further study.\footnote{ Such a changeover is generically observed on crystal:  In fact a recent skyrmion crystal calculation with hidden local symmetry Lagrangian -- without unknown parameters -- confirmed the topological change at low enough density~\cite{ma-crystal}. There is also an independent support coming from renormalization-group analysis at one-loop order for the changeover of the parameters exploited in this paper~\cite{PLRS}. That a topological phenomenon is involved suggests that it is likely robust.
The quark condensate $\la\bar{q}q\ra$ vanishes on average in unit
cell in the crystal description but this is not a bona-fide order
parameter since the pion is present in the system, indicating chiral
symmetry is not restored in the half-skyrmion state.  We defer details to a later publication.}  This occurrence could be largely due to the use of a
constant (non-scaling) nucleon mass beyond $n_{1/2}$ -- that we assume here--
as is indicated in the half-skyrmion matter and predicted in the
parity-doublet model with a large chiral-invariant mass $m_0$.
It may be pointed out  that y(n) plays an important role in determining
the EoS of nuclear matter in Region II as illustrated in Fig. 3.

For density  $n>2.0n_0$, the parameters used for
 the EoSs (A) and (B) there differ only in
y(n)=0.77 for (A) and = 0.78 for (B), the $c_{II}$
parameters used for them being identical. It is seen
from Fig.~3  that this small difference
 has made the  (A) EoS significantly more repulsive than (B),
 especially in the high density region.
Our calculations have found that  the use of a smaller
y(n)  would generally give  an upward lift to the
$n>n_{1/2}$ EoS, resulting in a  stiffer EoS.

We have  performed additional calculations studying mainly how our calculations depend on the location of the transition density $n_{1/2}$. Where it is located and how to choose is a central issue in our approach. It cannot be lower than $n_0$ since it will be at odds with nuclear structure as we know it. If it is far greater than $n_0$, then it will be inaccessible by terrestrial experiments, so will be difficult to verify its existence. As suggested in skyrmion crystal calculations~\cite{multifacet,LPR},  we will assume that it is located slightly above $n_0$.  Our results reported there indicate, however, that the  EoSs obtained with the new-BR scaling in the range $1.5n_0<n_{1/2}<2.0n_0$  depend only weakly on $n_{1/2}$ picked for all ranges of density relevant for our work. We believe that the precise location of $n_{1/2}$  is not important in our calculations as long as it is not far from $n_0$, while  it is $y(n)$ which plays an important role.

 By way of heavy-ion collision experiments,
there has been much progress  in determining
the nuclear symmetry energy $E_{sym}$ up to densities as high
as $\sim 5n_0$
\cite{li05,li08,tsang09}.
Thus an application of our new-BR  scaling
to the calculation of $E_{sym}$ would provide an important test
for this scaling in the region with $n>n_{1/2}$.

\begin{figure}[here]
\scalebox{0.34}{\includegraphics[angle=-90]{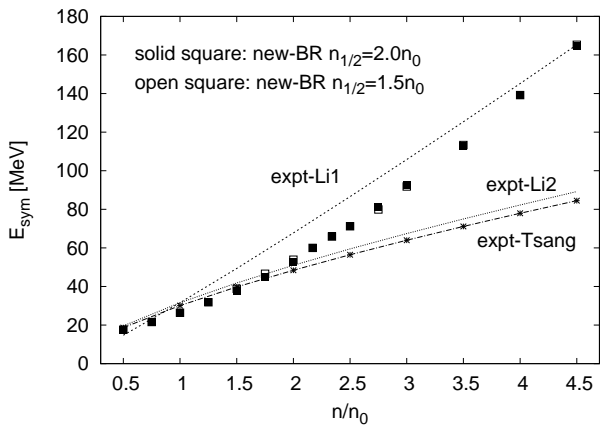}}
\caption{Comparison of our calculated
nuclear symmetry energies
with the empirical upper (expt-Li1)
and lower (expt-Li2) constraints
of Li et al. \cite{li05} and the empirical results of
Tsang et al. (expt-Tsang) \cite{tsang09}.
See text for more explanations.}\label{figure5}
\end{figure}

The nuclear symmetry energy
is related to the asymmetric nuclear matter EoS as Eq.~(\ref{E}).
We have calculated  $E(n,\alpha)$ for a range of $\alpha$
values, and  from them we extract $E_{sym}$.
Also we use the same ring-diagram formalism
where the $pphh$ ring diagrams are summed to all orders.
 In Fig.~\ref{figure5} we present our results  calculated with the
same new-BR A- and B-parameters mentioned earlier, labelled respectively
by solid- and open-squares there.
 It is of interest that the symmetry energies given by the A-
and B-parameters are nearly identical, despite the considerable differences
between the two corresponding EoSs for symmetric nuclear matter
shown in Fig. 3.

Based on heavy-ion scattering experiments, Li \emph{et al.} \cite{li05}
have suggested an empirical relation
\begin{equation}
E_{sym}(n) \approx 31.6(n/n_0)^\gamma;~ \gamma=0.69-1.1,
\end{equation}
for constraining the density dependence of the symmetry energy.
The upper ($\gamma = 1.1$) and lower ($\gamma=0.69$) constraints are also
plotted in the figure, labelled respectively as `expt-Li1'
and `expt-Li2'.  Also based on such experiments,
Tsang \emph{et al.} \cite{tsang09}
recently proposed a new empirical formula for the symmetry energy, namely
\begin{equation}
E_{sym}(n)=\frac{C_{s,k}}{2}\left(\frac{n}{n_0}\right)^{2/3}+
\frac{C_{s,p}}{2}\left(\frac{n}{n_0}\right)^{\gamma_i}
\end{equation}
where $C_{s,k}=25 {\rm MeV}$, $C_{s,p}=35.2 {\rm MeV}$
and  $\gamma_i \approx 0.7$. This formula is also plotted
in Fig.~\ref{figure3}, labelled as `expt-Tsang'. Note that Tsang's results are very
close to the lower constraint of Li et al.

\begin{figure}[here]
\scalebox{0.34}{\includegraphics[angle=-90]{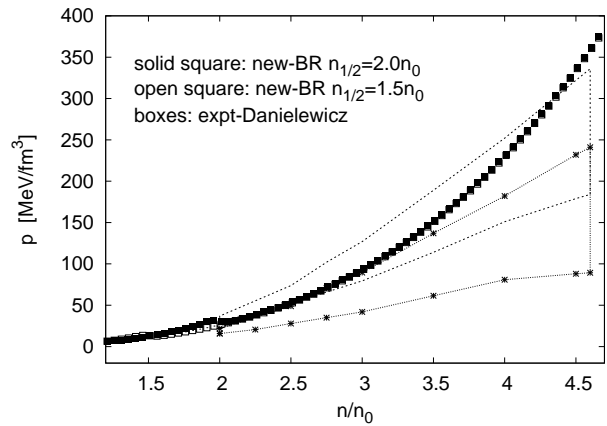}}
\caption{Comparison of our calculated neutron matter EoS
with the empirical stiff (upper box) and soft (lower box)
constraints of Danielewicz {\it et al.} \cite{daniel02}.}\label{figure6}
\end{figure}

Returning to Fig.~\ref{figure5}, we see that our new-BR results agree
reasonably well with the empirical constraints on $E_{sym}$; for Region I
our results are slightly below the empirical lower bounds while
in Region II they tend to be closer to the upper bound. This accounts for the EoS becoming stiffer over empirical fits as shown in Fig.~\ref{figure6} for neutron matter, where symmetry energy is active in its full strength with $\alpha=1$, while the EoS for symmetric matter (with no contribution from the symmetry energy, $\alpha=0$), lies within the empirical range as seen in Fig. ~6.

Recently Lattimer and Lim \cite{lattimer12}  have investigated the constraints on $E_{sym}$ and  $L$ (defined as $ 3u(dE_{sym}/du) , u\equiv n/n_0$)
at density $n=n_0$. The results deduced from nuclear masses, nuclear giant
dipole resonances, astrophysics, neutron skins of the $Sn$ isotopes, and
other investigations exhibit wide variations, with
$E_{sym}/MeV$ ranging from $\sim24$ to $\sim 36 $ and $ L/MeV$
from $\sim -20$ to $\sim 100$.
The overlap constraints allowed by all these results are
$29.6 \lsim E_{sym} /MeV\lsim 34 $ and $ 40.5\lsim L/MeV \lsim 61.9$ \cite{lattimer12}.
Our results as given by (A) and (B) of Fig.~4 are $E_{sym}/MeV \eqsim 26.5$
and $L/MeV \eqsim 64$. They are compatible with the respective bounds of the
above constraints, as well as consistent with the constraints given by the nuclear masses and nuclear giant dipole resonances.  We should of course emphasize that the result that our values for $E_{sym}$ and $L$ determined near the nuclear matter density lie at the border of  the bound does not directly reflect on the quality of our new-BR which brings in new ingredient in Region II following the topology change in our theory. In fact it concerns mainly the parameters of Region I which could be suitably readjusted to agree better with the bound without affecting other observables. In this regard, we are not in agreement with the currently favored notion that certain dense matter theories can be ruled out by the bounds.

\begin{figure}[here]
\scalebox{0.34}{\includegraphics[angle=-90]{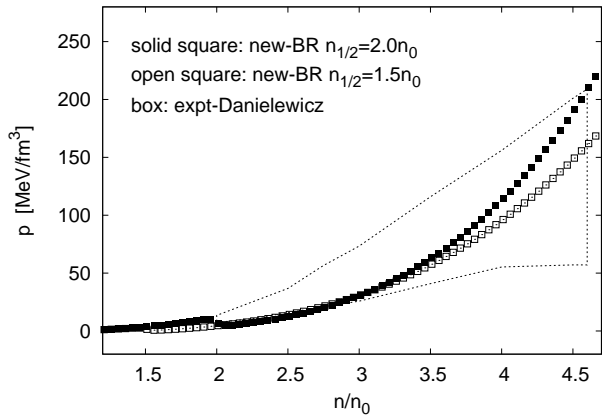}}
\caption{Same as Fig. 5 for symmetric nuclear matter.}\label{figure7}
\end{figure}
\begin{figure}[here]
\scalebox{0.34}{\includegraphics[angle=-90]{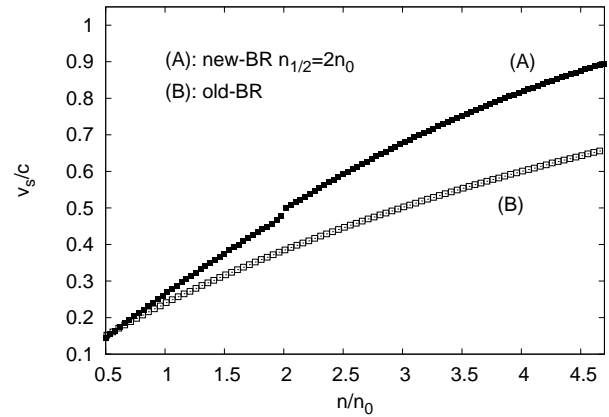}}
\caption{Speed of sound $v_s$ calculated with new- and old-BR
scalings. }\label{figure8}
\end{figure}

From heavy-ion collisions, Danielewicz {\it et al.} \cite{daniel02} have
obtained constraints for the pressure-density EoS $p(n)$
of neutron matter  up to  densities $\sim 4.5 n_0$.
To further study our new-BR scaling in the high density region,
we have calculated the neutron $p(n)$ EoSs up to the above densities.
A comparison of our results
with their constraints is presented in Fig.~\ref{figure6} where the upper
and lower boxes  are respectively the
constraints for the stiff and soft EoSs of \cite{daniel02}.
 Our EoSs calculated with parameters A and B are denoted by
 `solid-' and `open-square' respectively.  A similar comparison for
 the $p(n)$ EoSs for symmetric nuclear matter is  presented
in  Fig.~\ref{figure7}. We have also calculated the speed of
sound $v_s$ in nuclear (and neutron) matter using the relation
$(v_s/c)^2=dp/d\epsilon$, $\epsilon$ and $c$ being respectively
the energy density and speed of light. As an illustration,
our results for $v_s$ in neutron matter are presented in Fig.~\ref{figure8}. {The results are given for the range of density for which our theory is applicable. The extrapolation procedure used to go higher in density so as to obtain the maximum star mass is described below.}

As seen from Figs.~\ref{figure6} and \ref{figure7}, our calculated pressures are in satisfactory agreements with the empirical constraints of \cite{daniel02}.
 Note, however, our results are  somewhat stiffer than
the experimental ranges at high densities near $\sim 4n_0$. It is instructive to look at
the speed of sound $v_s$  which is closely related to the stiffness of the EoS.
It is seen from  Fig.!\ref{figure8} that the $v_s$ given by our new-BR scaling is significantly larger than that given by the old-BR scaling \cite{dong09}, indicating the former EoS
being stiffer. As to be reported below,
our  neutron-matter EoS is, however, somewhat stiffer
than what would give the 2-solar mass star. There seems to
 be nothing obviously wrong with this. The model having the
hybrid hadron-quark continuity mentioned above~\cite{masuda} seems to
favor such massive stars.

We have not yet described how we calculate the $p$ and $v_s$
 results shown in the above 3 figures. Let us do this now.
Including the nucleon rest-mass energy, we  first calculate the
nuclear-matter energy density
\begin{equation}
\epsilon (n)=n(\frac{E_0(n)}{A}+m_N)
\end{equation}
with the average ground-state energy
$E_0(n)/A$ obtained from the ring-diagram method described
earlier (section~3). The pressure-density EoS is then given by
\begin{equation}
p (n)=n\frac{d\epsilon(n)}{dn} - \epsilon(n).
\end{equation}

 As indicated above, to calculate  $p(n)$ we need to have
the derivatives of the energy EoS $d\epsilon (n)/dn$ or $d(E_0(n)/A)/dn$.
There is, however, a difficulty in doing so, as  our $E_0(n)/A$ EoS as shown
in Fig. 3   is 'not' a continuous/smooth one: It is composed of two branches,
one for skyrmion ($n<n_{1/2}$) and the other for half-skyrmion ($n>n_{1/2}$).
These two branches have clearly different shapes (slope
and curvature), and their slopes are not continuous at $n_{1/2}$.
Also the EoS after $n_{1/2}$  has a short segment of
 plateau-like structure at $n\eqsim 2.2n_0$.
These features present obstacles to the  calculation of
the derivatives  $d\epsilon (n)/dn$ and consequently hinder the
calculation of $p(n)$.
 To circumvent this difficulty, we need to employ a fitting procedure
so as to have a smooth (differentiable) $\epsilon(n)$.
Such a smooth crossover is expected also in a hybrid hadron-quark
matter model mentioned below~\cite{masuda}.

\begin{figure}[here]
\scalebox{0.34}{\includegraphics[angle=-90]{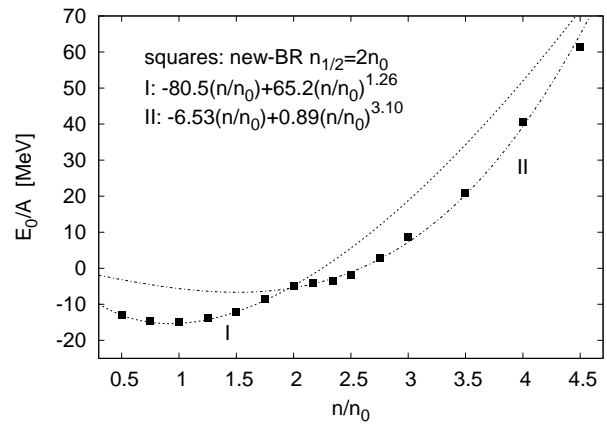}}
\caption{Polytrope fits of the $n_{1/2}=2n_0$ $E_0/A$
EoS of Fig. 3.}\label{figure9}
\end{figure}

 Li and Schulze \cite{schulze08} recently proposed a highly desirable
parametrization
for the nuclear-matter EoS: they have found that  a wide range of nuclear
EoSs can be  fitted very accurately by the polytrope EoS
$E_0(n)/A=a~ n + b~ n^{c}$ where $a$, $b$ and $c$ are
parameters.
 We have adopted this fitting procedure in our present work.
With such polytrope EoSs, the pressure EoS $p(n)$ can be
conveniently obtained and so is the speed of sound $v_s$
($(v_s/c)^2=dp/d\epsilon$).
 To illustrate this fitting, let us consider its application
 to the $n_{1/2}=2n_0$ EoS of Fig. 3. We have found it impossible to fit the
EoS entirely with one polytrope. But with two polytropes, one for skyrmion
and another one for half-skyrmion, a  satisfactory fit to the entire
EoS can be achieved as shown in Fig.~\ref{figure9}.
(In our fitting, we actually use the polytrope of the form
$E_0(n)/A=a~ (n/n_0)
 + b~ (n/n_0)^{c}$. In this way, the coefficients $a$ and $b$
 have the same units (MeV) and $c$ is dimensionless.)
As seen, the fit comes out quite well. Furthermore, the $a$, $b$ and $c$
coeficients for the two polytropes are vastly different. This is a
worth-noting result, suggesting
that the skyrmion and half-skyrmion EoSs are largely different
'mathematically'. Are they also very different physically?
It should be useful and of much interest to investigate this question
theoretically as well as experimentally.

\begin{figure}[here]
\scalebox{0.34}{\includegraphics[angle=-90]{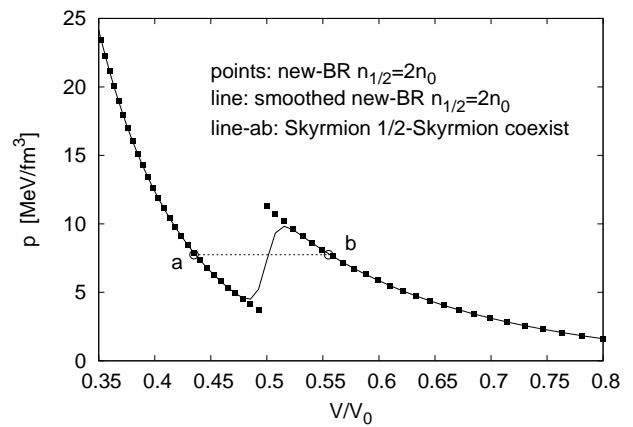}}
\caption{Skyrmion half-skyrmion coexistence
in symmetric nuclear matter calculated with  $n_{1/2}=2n_0$ new-BR
scaling. The coexistence points a and b satisfy both
$p(a)=p(b)$ and $\mu(a)=\mu(b)$.}\label{figure10}
\end{figure}

Our results for the $pV$ diagram originated from
 the $n_{1/2}=2n_0$ EoS of Fig. 3
are presented in Fig.~\ref{figure10}. Here the volume is defined as $V/V_0\equiv n_0/n$.
As seen, $p(n)$ is discontinuous at the cross-over density $n_{1/2}$.
Furthermore, at this point the half-skyrmion pressure
is significantly lower than the skyrmion pressure. This relative
difference in pressure is a necessary condition for  having a
skyrmion half-skyrmion coexistence. (The coexistence would not be possible
if this relative difference were reversed.) To have such a coexistence,
we also need to have the two coexistence points, labelled  $a$ and $b$
in the figure,
satisfying simultaneously pressure and chemical-potential equivalences,
namely $p(n_a)=p(n_b)$ and $\mu(n_a)=\mu(n_b)$. ($\mu=d\epsilon/dn$,
$\epsilon$ being the energy density.) The points $a$ and $b$ of Fig.~8
satisfy this double requirement, with $n_a=2.23n_0$, $n_b=1.72n_0$,
$p(n_a)=p(n_b)=~6.59 ~MeV/fm^3$ and $\mu_a=\mu_b=14.61 ~MeV$.

 The above results are for the $n_{1/2}=2n_0$ symmetric nuclear matter
 using the new-BR A-parameters. We have repeated this calculation for
neutron matter, obtained $(n_a,n_b)$=(1.93,~2.09)$n_0$. The width of the
coexistence region is about $0.15n_0$, considerably narrower than that
for symmetric nuclear matter. For the $n_{1/2}=1.5n_0$
calculation using the
B-parameters, we have obtained
$(n_a,n_b) $ = (1.49,~1.89)$n_0$ for symmetric nuclear matter,
and =(1.43,~1.59)$n_0$ for neutron matter. Note that here the $n_a$ for
the symmetric nuclear matter  is very close to the cross-over density
$n_{1/2}$ (=1.5$n_0$). This  suggests that the
 skyrmion-half-skyrmion transition in this case is almost a pure unison
cross-over where the nuclear matter at $n<n_{1/2}$ is entirely composed
of skyrmions, and when density increases to $n_{1/2}$ it all
becomes  half-skyrmion matter, leaving no
buffer zone for their coexistence.

 In Fig.~8, the `smoothed' $pV$ curve is obtained by combining
the two discontinuous branches  using two Fermi-Dirac functions,
similar to what we did in smoothly
joining the scaling functions $\Phi_I$ and $\Phi_{II}$  described
in section 3 (see (3.6) and (3.7)). The resulting $pV$ curve is
then of the standard
form for coexistence, like that for the familiar
liquid-gas coexistence. The above smoothing procedure has also been used
 for the pressure EoSs of Figs.~\ref{figure5} and \ref{figure6}.


\begin{figure}[here]
\scalebox{0.34}{\includegraphics[angle=-90]{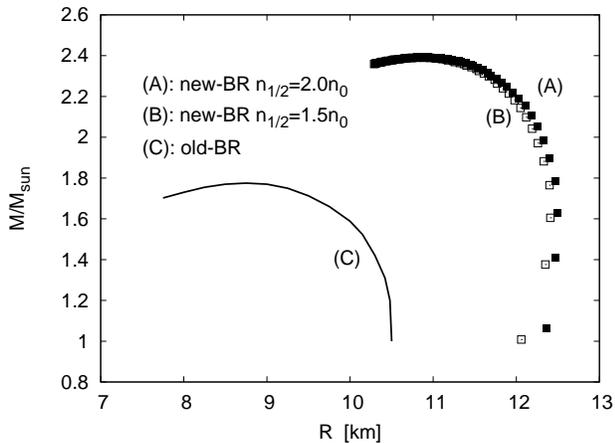}}
\caption{Mass-radius trajectories of
neutron stars calculated
with new-BR scalings using $n_{1/2}$ = 2.0 (A) and 1.5$n_0$ (B).
The maximum neutron-star
mass and its radius for these two cases are
respectively  (2.39 $M_{\odot}$, 10.90 km) and
(2.38 $M_{\odot}$,  10.89  km). The prediction with oldBR (C)~\cite{dong09} is given for comparison.}\label{figure11}
\end{figure}

\begin{figure}[here]
\scalebox{0.34}{\includegraphics[angle=-90]{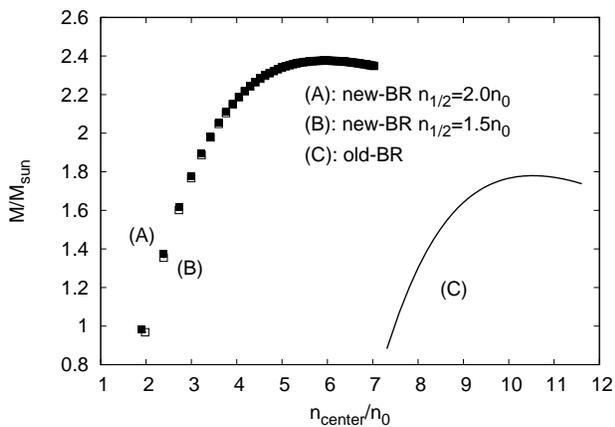}}
\caption{Central densities ($n_{center}$) for the neutron stars
of Fig. 9.}\label{figure12}
\end{figure}

In a recent neutron-star calculation using realistic NN potentials
\cite{dong09}, the effects from the `old-BR'
scaling (\ref{oldBR}) and (\ref{oldBR1}) applied in both
I and II were found to be highly important for neutron stars,
the maximum mass and its radius calculated (with, without) the
inclusion of such effects being respectively
($\sim 1.8, \sim 1.2 M_{\odot}$) and ($\sim 8.9, \sim 7.2$)km.
Now the question is: What does the new-BR (\ref{newBR}) do to
neutron stars? To address this question, we have calculated
 the properties of pure neutron stars (i.e. made of neutrons only)
from the above neutron-matter EoSs, using the calculation procedures
described in \cite{dong09}.

In  Fig.~\ref{figure11}  we present our calculated neutron-star mass-radius
trajectories (A) and (B),
obtained respectively with the A- and B-parameters mentioned earlier.
(In this figure the symbol $M_{sun}$ is used to denote the solar mass
$M_{\odot}$.)
The caveat mentioned above notwithstanding, it is interesting that the main results of the two calculations (A) and (B) are nearly the same.
The maximum mass of neutron stars given
by the two are  practically identical as given in the
caption of Figure \ref{figure11}. It is significant that there is little dependence on the location of $n_{1/2}$ as long as it is not too high above $n_0$. In the low-mass region (lower-right
corner of the figure) the trajectories are noticeably different, with
(A) having slightly larger mass and longer radius. These results
are consistent with the results shown in Fig. 5 where the two EoSs are
essentially equivalent to each other except in the narrow region
between 1.5 and 2.0$n_0$.
The effects from the present
new-BR scaling appear to be even stronger than those from the
old-BR scaling. For example, the maximum mass obtained (with, without)
the  new-BR scaling are ($\sim 2.4, \sim 1.2 M_{\odot}$), the
increase between them being
significantly larger than the above old-BR case. As one can see in Fig.~\ref{figure11} (see also Fig.~\ref{figure12}), the star properties are markedly different between the old-BR and the new-BR.
By analysing a wide range of empirical data, Steiner et al.
 \cite{steiner2010} have obtained a constraint for neutron-star radius
 $ 10km \leq R \leq 12.5km$. The radius given
 by our new-BR calculations is in good agreement with this
 constraint.

In Fig.~\ref{figure12} we report  the central
densities $n_{center}$ of the neutron stars calculated with the
new-BR scalings as discribed above. The maximum-mass neutron
stars have $n_{center}\simeq 5.5n_0$.
Recalling Figs. 5 and 6, Danielewicz
{\it et al.} \cite{daniel02} have provided experimental constraints
for the nuclear-matter EoSs up to density $\sim 4.5n_0$.
This gives us
important guidelines about the EoSs below this density. But beyond this,
there is still no such guidelines and one is really
not at all sure what the EoSs there should be.
Although we can 'calculate' the EoS
using our new-BR formalism up to any densities, the resulting EoS
is, we believe, of 'good confidence level' only for densities
below and
not much higher than $\sim4.5 n_0$. Thus we have adopted
an extrapolation scheme, namely
calculating the EoS up to an extrapolation density $n_{ext}$  while
obtaining the EoS beyond this density
by a polytrope extrapolation.
(The polytrope is obtained by fitting
the EoS below $n_{ext}$.) { The mass of neutron star with central density of $n_{ext}$
is $\sim 2.3M_{\odot}$. }

Clearly this extrapolation can be applied
only to densities not too much higher than $n_{ext}$.
 We have employed
$n_{ext}$= 4.5 and 5.5$n_0$ and found that the EoSs given by them
are  in close agreement with each other  up  to  $\sim 7.0n_0$.
This and that our $n_{central}$ is as small as  $\sim 5.5n_0$
    support the reliability of
 the above extrapolation procedure for our present neutron-star calculations. {The causal limit in this extrapolation appears at $5.9n_0$, which is larger than the  central density of $5.5n_0$ for the maximum mass in Figs. 11 and 12. }
 It would be very interesting to study the EoS
for neutron stars with RIB machines
as the low central densities of neutron stars as given earlier should
be readily accessible there.
 As stated earlier, our present calculation has assumed a
pure-neutron-matter composition for neutron stars without
taking into account a variety of compact star conditions. This could be
an oversimplification, and
 the results obtained thereby should be taken, at best,
indicative of what could be happening in nature.

\section{Comments and Discussions}
In this paper, we subjected the nuclear effective field theory anchored on RG flow, with the parameters of the Lagrangian sliding with density, to normal nuclear matter and dense compact-star matter. The scaling behavior used here differs from the old  BR scaling~\cite{BR91}, in that at a density $n_{1/2}>n_0$, a topological change takes place from skyrmion matter to half-skyrmion matter,  giving rise to a modified scaling  new-BR. The changeover from skyrmion matter to half-skyrmion matter is characterized by a vanishing quark condensate $\la\bar{q}q\ra=0$ but a nonvanishing pion decay constat $f_\pi\neq 0$. Thus it is not a standard phase transition \`a la Ginzburg-Landau-Wilson paradigm although two different phases are involved; it appears to involve an emergent symmetry not present in the fundamental theory, QCD.

At the semi-classical approximation made in the calculation, the half-skrymions are not deconfined in contrast to what happens in certain condensed matter systems~\cite{deconfined}. They are bound or confined, so they are not propagating degrees of freedom. What characterizes the system is that the mass of the baryon made up of two `bound' half-skyrmions remains more or less unscaled, not going to zero up to the density $n_c$ at which the quarks get deconfined, whereas the $\rho$-meson mass is expected to drop faster in the half-skyrmion phase than in the skyrmion phase. This means that the origin of the most,  if not all, of the nucleon mass is not in the dynamical symmetry breaking of chiral symmetry, in contrast to the meson mass,  with a substantial mass of the nucleon coming from a hitherto unknown source. This is similar to what is described in the parity-doublet model of the nucleon~\cite{detar,PLRS}.We should note however that this picture is clearly at odds with the constituent quark model -- which has a strong theoretical support from QCD in the large $N_c$ limit~\cite{weinberg} -- where the ratio of the meson mass over the baryon mass is  2/3. Whether or not the constituent quark model is applicable in nuclear medium is not known, but if there were an $m_0$ for the quark which is not small, then it should be possible that  the constituent quark model hold in dense medium and the ratio remain more or less the same. In this case, the scaling could be considerably different from the new-BR.

It is intriguing that the two consequences of the changeover at $n_{1/2}$, namely, the drastic modification of the nuclear tensor force and the stiffening of the EoS of dense matter at $n_{1/2}$, seem to be hinting at the mechanism for the generation of  $\sim 99$ \% of the nucleon mass in the strong interactions.  See \cite{MR-mass} for discussions on this matter.

The salient features obtained in the RG-implemented effective theory approach adopted in this paper can be summarized as follows:
\begin{enumerate}
\item  Without a suitable scaling in the Lagrangian that
figures in $V_{low-k}$ (or incorporating many-body forces),
symmetric nuclear matter cannot be stabilized at the right
density and with correct binding energy.
\item Our calculations have essentially
two scaling parameters:
one is $c_I \approx 0.13$ for all mesons (vector mesons and
scalar meson) and the nucleon in region I, and in region II
 we have $c_{II}=c_{I}$ for mesons and the vector coupling and
an additional parameter $y(n)\approx 0.8$ for
the nucleon. With these two parameters, one can explain
satisfactorily the saturation density, the binding energy and the
compression modulus of symmetric nuclear matter as well as the nuclear
symmetry energy, and predict the EoSs
for symmetric and asymmetric nuclear matter at high density and
compact-star matter. Our results give a good
fit to all quantities that are available experimentally at densities
up to $n\sim 4n_0$.

\item The topology change from skyrmion to half-skyrmion at $n_{1/2}$
changes the slope of the EoS, making it stiffer in the half-skyrmion
phase and raises the maximum mass of compact stars to $\sim 2.4M_\odot$.
 Verifying the presence and the role of the topology change
at $n_{1/2}$ should be feasible at RIB machines.

\end{enumerate}

 In our treatment, $n$-body forces for $n>2$ have not been taken into account. As mentioned, 3-body forces -- in place of BR -- could equally well provide the repulsion needed to stabilize nuclear matter. This does not mean that the many-body forces and the BR are alternatives. They should both come in together. In principle, there should be no problem in including both BR and many-body forces in a way consistent with the tenet of chiral expansion. What one has to do in the presence of such n-body potentials is then to suitably modify the scaling properties of the Lagrangian, since direct and indirect chiral symmetry effects are compounded in physical quantities in a variety of different chiral expansion schemes as illustrated in \cite{FR}. A fully consistent way of doing the calculation would be to have {\em both} the scaling and many-body potentials treated together with certain constraints, such as thermodynamic consistency, taken into account.  We also note that our EoS is very close to the EoS found in Ref.[40] with a similar stiffening throughout the range of density considered, where  the sound velocity never exceeds 0.9.

We have not taken into account strangeness degrees of freedom -- such as kaons, hyperons, strange quarks etc. -- into the EoS for neutron-rich matter. In our formulation anchored on dense skyrmion matter, as described in \cite{LR12}, hyperons can enter only {\em after} kaons condense. Therefore the issue here is how kaon condensation can take place after changing from skyrmion matter to half-skyrmion matter.

There are two opposing mechanisms to consider in the process. One is that in the presence of the topology change at $n_{1/2}$, the mass of $K^-$ has a propitious drop not present in conventional chiral perturbation treatments~\cite{PKR}. This goes in the direction of lowering the critical density for kaon condensation. The other is the effect of stiffening the EoS.  It is known for instance in phenomenological studies that the more repulsion  there is in non-strange nuclear interactions, the higher the kaon condensation critical density goes up~\cite{pandha}.  What will happen in compact stars therefore will depend crucially on which one dominates. One intriguing possibility is that the stiffening postpones the drop of $m_K^*$ in a manner analogous to the stiffening at the smooth crossover at a density $\sim (2-4) n_0$ from hadron to non-strange quark phase in the hybrid model that also yields the maximum star mass $\sim 2.3 M_{\odot}$~\cite{masuda}. This will also have an important impact on the cooling of the star, since the appearance of strange flavor at higher density will prevent fast direct URCA process from setting in too precociously.

 It should be stressed that in our approach, strangeness in the form of condensed kaons (or equivalently hyperons) may enter at near or even before the density to which our theory with topology change can be extended, say $\sim 4.5n_0$. Therefore the extrapolation beyond such density with polytropes, without accounting for strangeness degrees of freedom,  potentially violating causality, should be  taken as merely exploratory.

One important aspect in our treatment that requires serious studies is the correlation between the behavior of the in-medium nucleon mass $m_N^*$ and  that of the in-medium $\omega$-N coupling $g_{\omega NN}$ which is related to the $U(1)$ gauge coupling $g^*_\omega$. We have adopted in our calculation the information from the skyrmion crystal calculations~\cite{half,ma-crystal} and the parity-doubling nucleon model~\cite{PLRS} that the nucleon mass drops only about 20\% up to the highest density we are considering.  We have taken the scaling $y$ effective in Region II to be constant as indicated in the skyrmion-crystal calculation~\cite{ma-crystal} and in the one-loop RG analysis of HLS Lagrangian. As stated, were we to drop the $\omega$-nucleon coupling according to $g^*_\omega/g_\omega\approx g^*_\rho/g_\rho\approx g^*/g=\Phi_{II}$ as one would expect if flavor $U(2)$ symmetry held in Region II, the EoS would become much too soft above $n_0$ to be compatible with the existence of the 2-solar mass object observed in nature. We kept $g^*_\omega/g_\omega\approx 1$ while letting the $\omega$ mass scale.
Now to quantify the above observation, we have examined
the effect of dropping $\omega$-NN coupling for
given $m_N^*$s. Writing the $\omega$-nucleon coupling
in Region-II as $g^*_\omega/g_\omega=(1+c_{II,N\omega}~n/n_0)^{-1}$,
we have found at  $n=2.5n_0$, $E_0/A= (-33.9, -50.9,-68.4) $ MeV
for y(n)=0.77 and $E_0/A= (11.68, -1.26, -14.56)$ MeV for
y(n)=0.60 for the scaling constant of the $\omega$-NN coupling
$c_{II,N\omega} = (0.046, 0.093, 0.139)$ with all other parameters
fixed to (A) of Fig. 3. One sees that the EoS is extremely sensitive
to the in-medium properties of both the nucleon mass
and the $\omega$-NN coupling.

There are two implications that follow from this calculation. One is that $U(2)$ symmetry can be badly broken in dense medium and as a consequence the vector manifestation of HLS~\cite{HY:PR} does not apply to the in-medium $\omega$ meson although its mass may approach zero as the $\rho$ mass does \`a la mended symmetry. The other is that the in-medium nucleon mass and $\omega$-NN coupling must be strongly correlated.  One-loop renormalization group equations with the generalized hidden local symmetry Lagrangian implemented with baryons (with no dilatons) of \cite{PLRS1} show that in the chiral limit, both $m_\rho^*$ and $m_\omega^*$ approach zero as the dilaton limit fixed point is approached. So does the nucleon mass $m_N^*$ in the standard (or ``naive") assignment for the nucleon (see \cite{PLRS1}). However while the vector manifestation of HLS~\cite{HY:PR} requires that $g_\rho^*/g_\rho\propto \la\bar{q}q\ra^*/\la\bar{q}q\ra\rightarrow 0$ near chiral restoration, if $U(2)$ symmetry is violated in medium, the in-medium $\omega$-NN is predicted to drop much more slowly than the $\rho$-NN coupling~\cite{PLRS}.  At one-loop order the $\omega$-nucleon coupling is found not to scale. It is only at two-loop and higher order that scaling sets in. One can see from the RGEs the interplay between the slow scalings of the coupling and nucleon mass.   This behavior agrees qualitatively with what was noticed above where lowering the nucleon mass required reducing the coupling $g_\omega$ in order to have the symmetry energy lie within the range given by heavy-ion data.

\subsection*{Acknowledgments}
Two of us (HKL and MR) are grateful for discussions with Masayasu Harada, Won-Gi Paeng and  Chihiro Sasaki. This paper was completed during the 2013 WCU-APCTP Focus Program at APCTP, Pohang, Korea. The work reported here was partially supported by the WCU project of Korean Ministry of Education, Science and Technology (R33-2008-000-10087-0),  the US Department of Energy under Grant No. DE-FG02-88ER40388 and DE-FG02-03ER41270, and the US National Science Foundation under Grant No. PHY-0099444.

\vskip 0.3cm

\end{document}